\documentclass[prb, preprintnumbers,amsmath,amssymb, superscriptaddress, twocolumn]{revtex4} %endfloats

\usepackage{graphicx}% Include figure files
\usepackage{dcolumn}% Align table columns on decimal point
\usepackage{bm}% bold math

\usepackage{color,ulem}

\begin{document}
\title{Broad-band coherent backscattering spectroscopy of the interplay between order and disorder in 3D opal photonic crystals.}
\author{Otto L. Muskens}
\affiliation{School of Physics and Astronomy, University of
Southampton, UK.}\email{O.Muskens@soton.ac.uk} \affiliation{Center
for Nanophotonics, FOM Institute for Atomic and Molecular Physics
AMOLF, Science Park 104, 1098XG Amsterdam, The Netherlands.}
\author{A. Femius Koenderink}
\affiliation{Center for Nanophotonics, FOM Institute for Atomic
and Molecular Physics AMOLF, Science Park 104, 1098XG Amsterdam, The
Netherlands.}
\author{Willem L. Vos}
\affiliation{Complex Photonic Systems, MESA
Institute for Nanotechnology, University of Twente, 7500 AE
Enschede, The Netherlands.}
\date{\today}

\begin{abstract}
We present an investigation of coherent backscattering of light
that is multiple scattered by a photonic crystal by using a
broad-band technique. The results significantly extend on previous
backscattering measurements on photonic crystals by simultaneously
accessing a large frequency and angular range. Backscatter cones
around the stop gap are successfully modelled with diffusion
theory for a random medium. Strong variations of the apparent mean
free path and the cone enhancement are observed around the stop
band. The variations of the mean free path are described by a
semi-empirical three-gap model including band structure effects on
the internal reflection and penetration depth. A good match
between theory and experiment is obtained without the need of
additional contributions of group velocity or density of states.
We argue that the cone enhancement reveals additional information
on directional transport properties that are otherwise averaged
out in diffuse multiple scattering.
\end{abstract}
 %}%} \narrowtext
\maketitle

\section{Introduction} Photonic crystals are ordered nanostructures with a
periodicity on the scale of the optical wavelength. The perfect
periodic ordering gives rise to a photonic band structure and
properties such as slow light, negative refraction, and photonic
band gaps.\cite{Joannopoulos} Photonic crystals are of importance
for their potential in controlling the transport and emission of
light at the nanoscale.\cite{Yab87,John87} Apart from the
properties of perfect photonic crystals, such as a 3D photonic
bandgap, the possibility of combining band structure with
controlled random multiple scattering has raised considerable
interest. Macroscopic photonic crystals can be assembled bottom-up
using colloidal suspensions, or top-down using lithographic
techniques. All crystals have some degree of intrinsic disorder
due to polydispersity of the building blocks or imperfections in
fabrication.\cite{Li2000,Vlasov2000,KoenderinkPLA,HuangPRL01,Astratov2002,Noda2002,KoenderinkPRL03,Allard2004,Rengarajan2005}
In order to realize random multiple scattering without relying on
intrinsic disorder, another route has been explored to
intentionally design disorder in three-dimensional photonic
crystals of very high quality.\cite{Garcia09} The role of disorder
has recently been investigated in photonic crystal
waveguides.\cite{Kuipers08, Topolancik07, Lodahl10}

The presence of disorder results in diffusive scattering of light
and a concomitant disappearance of the photonic band structure.
However, theoretical and experimental studies indicate that an
intermediate regime exists, where photonic band structure is
preserved locally but medium-range disorder disrupts ballistic
transport\cite{John87,KoenderinkPRL03,Kuipers08,Toninelli08,Erementchouk09,Garcia09}.
This intermediate scattering regime is reached when the scattering
mean free path $\ell$ exceeds the Bragg length\cite{remark1} $L_B$
but is smaller than the sample size $L$: $L_B<\ell<L$. As was
pointed out by Sajeev John, the modified density of states
associated with the photonic bands near a 3D photonic band gap may
result in large corrections to the diffuse transport, as is known
for completely random systems.\cite{John87} Recently it has been
argued that the scattering mean free path in opals with designed
disorder yield contributions from both the local density of states
and group velocity.\cite{Garcia09} Ultimately, the photonic
crystal could facilitate the complete breakdown of light diffusion
known as Anderson localization. Indeed, indications of Anderson
localization have been observed in recent studies on
one-dimensional photonic crystal
waveguides\cite{Topolancik07,Lodahl10} and nonlinear lattices
\cite{Segev07}. Despite recent experimental
efforts\cite{Toninelli08}, this regime has not yet been observed
in a three dimensional photonic crystal.

Effects of multiple light scattering in photonic crystals have
been studied before using a variety of reflection and transmission
measurements, including coherent backscattering of light
[Fig.~\ref{fig:setup}(a)]. Coherent backscattering (CBS) is a
multiple scattering phenomenon caused by the constructive
interference of time-reversed light paths in the medium in a cone
around the backscattering direction. Coherent backscattering has
been used to investigate multiple scattering of photonic crystals
at a few discrete wavelengths.\cite{KoenderinkPLA,
HuangPRL01,pursiainen07} Recently the first broad-band
investigations of opal photonic crystals around the stop band have
been reported.\cite{Baumberg09} Modifications of CBS cones around
the photonic stop bands have been observed which were attributed
to two contributions related to the modification of the diffuse
source by Bragg reflection, and to variations of the internal
reflection conditions at the escape angle of the diffuse light, as
shown schematically in Fig.~\ref{fig:setup}(b). Both effects can
significantly affect the path length distribution for multiple
scattering.

Here, we present experiments demonstrating the effect of the
photonic band structure on diffuse light scattering over a wide
spectral and angular range in polystyrene opals. We employ a
broad-band coherent backscattering technique using a
supercontinuum white-light source.\cite{muskensOE08} The technique
has been used successfully to investigate strongly photonic random
media, i.e. TiO$_2$ powder and porous GaP, and to characterize
resonant light trapping by semiconductor
nanowires.\cite{muskensNL09} Compared to earlier studies on
photonic crystals \cite{KoenderinkPLA,
HuangPRL01,pursiainen07,Baumberg09}, we achieve a more detailed
characterization of CBS as a function of both angle of incidence
and wavelength. This investigation over an extensive range of
angles and frequencies allows us to compare the results with
semi-empirical model calculations to verify earlier
predictions.\cite{KoenderinkPLA}

\section{Experimental methods}

\begin{figure}[t] \newpage
\includegraphics[width=8.2cm]{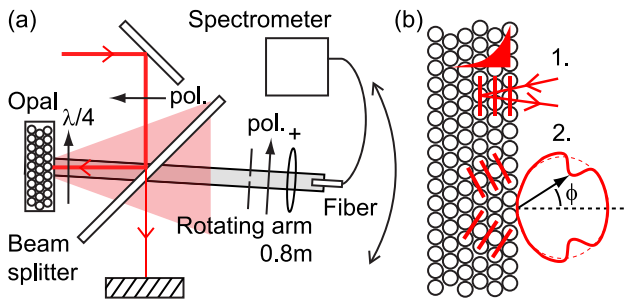}
\caption{\label{fig:setup}(color online) (a) Experimental setup
for coherent backscattering spectroscopy. (b) Photonic crystal
effects influencing the apparent mean free path: 1. modification
of source of diffuse light by Bragg diffraction, 2. modification
of angle-dependent escape probability due to internal reflection.
}
\end{figure}

\begin{figure}[t] \newpage
\includegraphics[width=8.6cm]{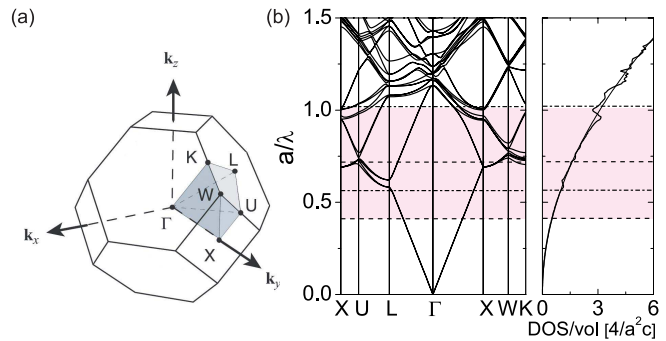}
\caption{\label{fig:reflspectra}(a) Wigner-Seitz cell of an
fcc-lattice with its irreducible part, showing symmetry points
$\Gamma$, \textit{L}, \textit{X}, \textit{U}, \textit{K},
\textit{W}. (b) Calculated band structure for a polystyrene opal,
with total density of states (DOS) of photonic crystal (line)
compared to effective medium (dash). Shaded area indicates the
reduced frequency range accessed in our experiments, with
horizontal lines indicating individual ranges of the 130-nm (dash)
and 180-nm (dash-dot) opals}.
\end{figure}

Coherent backscattering spectra were obtained over a wide spectral
range in the visible and near-infrared using a supercontinuum
white-light backscattering setup described previously in
Ref.~\onlinecite{muskensOE08}. The angular resolution was improved
to 0.8~mrad full width at half maximum, in order to resolve the
narrow CBS cones of photonic crystals.\cite{KoenderinkPLA}
Small-angle coherent backscattering measurements were obtained in
a beamsplitter configuration as shown in Fig.~\ref{fig:setup}(a).
Circular polarization channels were selected using achromatic
polarizers and a quarter wave plate directly in front of the
sample, with the purpose of reducing the contribution of single
scattering in the CBS enhancement factor. The opals used in this
study were obtained by sedimentation of polystyrene spheres (Duke
Scientific, polydispersity $\sim$2\%) in 0.3~mm thick, 3~mm wide
flat glass capillaries (Vitro Dynamics). We used two opals
composed of spheres of radii 130~nm and 180~nm, which were grown
from a suspension in water using a centrifuge.\cite{Vos96,Vos97}
After crystallization the water was slowly evaporated from the
capillaries to produce polycrystalline opals of several
millimeters in length. The polystyrene opal is an
\textit{fcc}-crystal with the [111] crystallographic direction
(corresponding to $\Gamma$\textit{L}) oriented perpendicular to
its surface, as was demonstrated by small-angle x-ray
scattering.\cite{Vos97} The corresponding Wigner-Seitz cell is
shown in Fig.~\ref{fig:reflspectra}(a), while
Fig.~\ref{fig:reflspectra}(b) shows the calculated band structure
and density of states. In our experiments we investigate the
behavior around the stopgap at L, which is located at
$\phi=0^\circ$ angle of incidence and has a bandwidth of 6\% as
predicted by band structure calculations and measured in many
reports on opals
elsewhere.\cite{Vos96,Romanov01,Galisteo2003,Vos01,Rengarajan2005}
By tilting the sample, we gain access to directions away from the
L-gap corresponding to \textit{LK}, \textit{LU}, and \textit{LW}
in the band diagram. In the experiments we average over many
crystalline regions with random orientations in the (111) plane,
and the crystal is spun around its $\Gamma$L-axis to average
individual speckle. Therefore all directions in the (111) plane
will contribute to our experimental results. Since the \textit{LK}
and \textit{LU} directions are very similar, we focus on the
\textit{LU} and \textit{LW} bands in the interpretation of our
results. Also shown in Fig.~\ref{fig:reflspectra}(b) is the
photonic density of states (DOS) (line), compared to that of the
homogeneous effective medium (dashed line). Clearly, density of
states variations amount to only several percent for opal photonic
crystals.\cite{Li01, Nikolaev08}

\section{Results}

\begin{figure}[t] \newpage
\includegraphics[width=8.7cm]{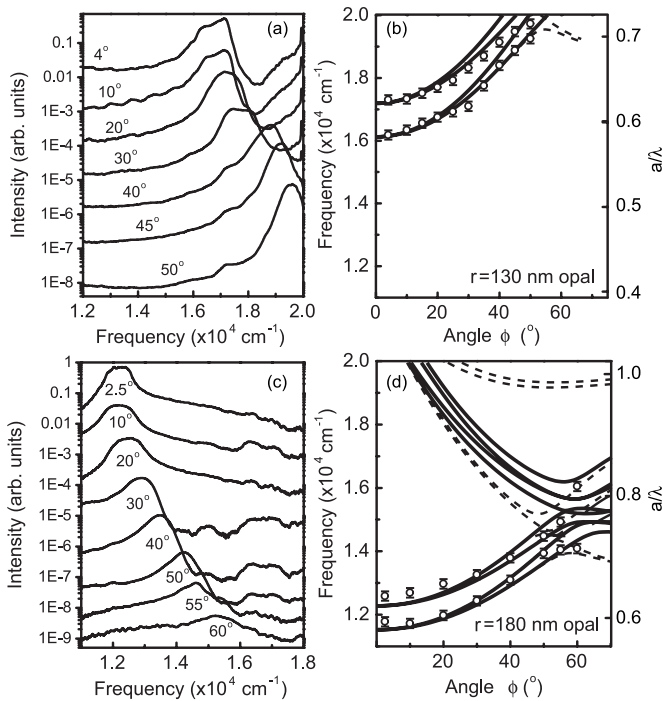}
\caption{\label{fig:reflectivities} Color online. (a) Experimental
reflectivity spectra of a $r=130$~nm opal for angles of incidence
from $\phi=4^\circ$ to $50^\circ$. Subsequent curves scaled by a
decade for visibility.(b) Dispersion relations for $r=130$~nm
opal, (circles) half-maximum values of the reflectivity peaks of
(a), (lines) calculations for fcc crystal along \textit{LU}
(dashed lines) and \textit{LW} (full lines) ($a=362$~nm). (c,d)
Same for $r=180$~nm opal (calculation, $a=510$~nm).}
\end{figure}

Specular reflectivity measurements were collected for different
angles of incidence $\phi$. Resulting spectra are shown in
Fig.~\ref{fig:reflectivities}(a) and (c) for the $r=130$~nm and
$r=180$~nm opals, respectively. For both opals a clear dispersion
is observed of the stop band, which is centered at 16800~cm$^{-1}$
for the $r=130$~nm opal and at 12200~cm$^{-1}$ for the $r=180$~nm
opal at near-perpendicular incidence. The frequencies of the
half-maximum band edges are shown in
Figs.~\ref{fig:reflectivities}(b) and (d) together with band
structure calculations, where the lattice constant was adjusted to
match the experimental stop band position at $\phi=0^\circ$. This
results in the dimensionless scaled units $a/\lambda$ as shown in
the right axes of Fig.~\ref{fig:reflectivities}(b) and (d), and
values of the lattice constant of 362~nm and 510~nm, respectively.
These values are in good accord with expected values based on the
sphere radii using $a\simeq 2 \sqrt{2}r$. The experimental
dispersion of the stop band matches well the expected behavior for
an \textit{fcc} opal. The full width at half maximum of the
reflection peaks of $\Delta \omega/\omega$ is in excellent
agreement with the photonic interaction strength $S=0.06$ defined
as the relative frequency width of the stop gap at the L-point
predicted by band structure calculations. In addition, higher
order bands approach the stop band at angles $\phi$ around
60$^\circ$, resulting in a more complex behavior of the reflected
intensity\cite{VanDiel2000, Romanov01, Lopez2004,Pavarini2005}, as
is observed for the $r=180$~nm sample. Given the spectral range
covered by our experimental setup, a set of samples is needed to
span nearly a factor three in frequency ($a/\lambda \sim 0.35 -
1.05$) that ranges from well below the first order L-gap to the
range of second order gaps that are the precursors of the inverse
opal band gaps.

\begin{figure}[t] \newpage
\includegraphics[width=8.7cm]{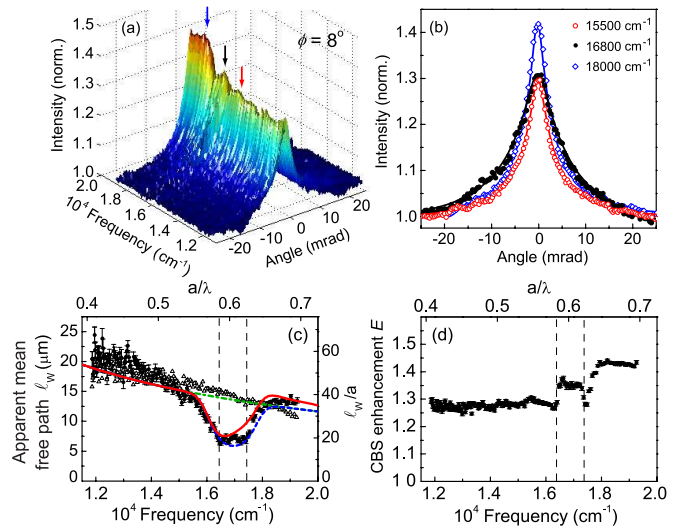}
\caption{\label{fig:CBSimage} Color online. (a) Experimental CBS
spectrum of a $r=130$~nm diameter opal at $\phi=8^\circ$ and (b)
Cross sections at frequencies of 15500~cm$^{-1}$ (circles, red),
17000~cm$^{-1}$ (dots, black), and 18000~cm$^{-1}$ (diamonds,
blue), with (lines) fits using diffuse scattering model. (c,d)
Values of apparent mean free path $\ell_W$ (dots) (c) and CBS enhancement
$E$ (d) obtained from fits to (a). Lines denote $\omega^{-0.96}$ dependence (dash-dot,green), obtained from experimental data at $\phi=50^{\circ}$ (open triangles), and calculations using Eq.~(\ref{eq:ellW}) with (line, red) and without (dash, blue) internal reflection correction.}
\end{figure}

Coherent backscattering spectra were measured for the $r=130$~nm
opal for various angles of incidence $\phi$. A typical CBS
spectrum taken at $\phi=8^\circ$ is shown in
Fig.~\ref{fig:CBSimage}(a). The CBS spectrum shows a pronounced
structure around 16800~cm$^{-1}$, corresponding to the center
frequency of the stop gap. The arrows indicate frequencies below,
inside, and above the stop gap, for which cross-sections are shown
in Fig~\ref{fig:CBSimage}(b). We note that the height of the CBS
cones is not at all related to the peak of the reflectivity in
Fig.~\ref{fig:reflectivities}(a). The traces show the typical CBS
lineshape, which can be fitted to a model of random light
scattering [lines in Fig~\ref{fig:CBSimage}(b)]. We employed a
finite slab model analogous to Ref.~\onlinecite{Mark88}, where the
slab thickness was fixed for the full data set. The residuals of
all fits are less than 5\%, illustrating a very good match. This
agreement indicates that the diffusion Green functions are
sufficient to calculate the path length distribution underlying
the CBS cones, without requiring the addition of photonic band
structure effects. The validity of the random scattering model is
consistent with earlier experiments.\cite{KoenderinkPLA} Theory
has indeed confirmed that isotropic diffuse transport is obtained
for crystal structures with high symmetry, even when scattering
between individual modes is highly
anisotropic.\cite{Erementchouk09}

 The model fits using the random scattering theory yield values for
the apparent mean free path $\ell_W$, which relates to the width
of the CBS cones according to $\theta_{\rm FWHM} \simeq
0.7/k_0\ell_W$, and the CBS enhancement factor $E$. Spectra of
both the apparent mean free path $\ell_w$ and the enhancement
factor $E$ (obtained from data in Fig.~\ref{fig:CBSimage}(a)) are
shown in Fig~\ref{fig:CBSimage}(c,d) (dots). The dashed vertical
lines indicate the high and low-energy edges of the photonic stop
band obtained from the reflectivity data. For comparison we also
show the $\ell_W$ spectra obtained at $\phi=50^\circ$ (open
triangles). At this angle, the photonic stop band has shifted
outside of the spectral range, and the frequency variation of the
mean free path mainly corresponds to a background resulting from
structural variations in the crystal.\cite{KoenderinkPRB05} We
modeled this variation of the mean free path to a power law
$\ell\propto \omega^{-\alpha}$, yielding an exponent
$\alpha=0.93\pm 0.04$ (dash-dotted line, green). This exponent is
much less than that of pure Rayleigh scattering ($\alpha=4$), and
also less than the Rayleigh-Gans dependence ($\alpha=2$) found for
the transmission of similar
opals.\cite{HuangPRL01,KoenderinkPRB05,Garcia09,VlasovPRB99,Tarban,Pradhan,Koerdt,Miguez,Park}
The low-frequency value of $\ell$ of around 20~$\mu$m translates
to a ratio $\ell/a \simeq 55$ for our crystal ($a = 362$~nm), in
agreement with the long-wavelength values observed for similar
samples in Ref.~\onlinecite{KoenderinkPRB05}.

For the $\phi = 8^\circ$ spectrum in Fig.~\ref{fig:CBSimage}(c), a
pronounced trough is observed in $\ell_W$ at frequencies that
correspond to the measured L-gap. A similar behavior has been
observed in experiments \cite{HuangPRL01, Baumberg09}, and was
first predicted in Ref.~\onlinecite{KoenderinkPLA} based on a
reduction of the penetration depth of the diffuse source by Bragg
reflection. Qualitatively, a reduced penetration depth implies
shorter paths, as paths are more likely to meet the sample
boundary and exit. Hence Bragg reflection causes cone broadening.
The reduction of $\ell_W$ extends far outside the stop gap,
especially toward the lower-frequency side, consistent with the
behavior of the reflectivity spectra in
Fig.~\ref{fig:reflectivities}(a). This observation shows the
importance of measuring over a large bandwidth in frequency,
including a significant range well below the stop gap: it is only
far below the stop gap that one retrieves a transport mean free
path that is a true measure of the disorder in photonic crystals.
The Bragg length associated with the L-gap can be estimated from
the interaction strength according to $L_B=2d_{111}/\pi S$,
yielding $L_B \simeq 4$~$\mu$m. This corresponds to a ratio
$\ell/L_B\simeq 3.8$, in the L-gap, which fulfills the inequality
for the intermediate scattering regime ($L_B<\ell<L$). The solid
and dashed lines in Fig.~\ref{fig:CBSimage}(c) represent model
fits which will be discussed further below in
Sec.~\ref{sec:comparison}.

\begin{figure} \newpage
\includegraphics[width=8.6cm]{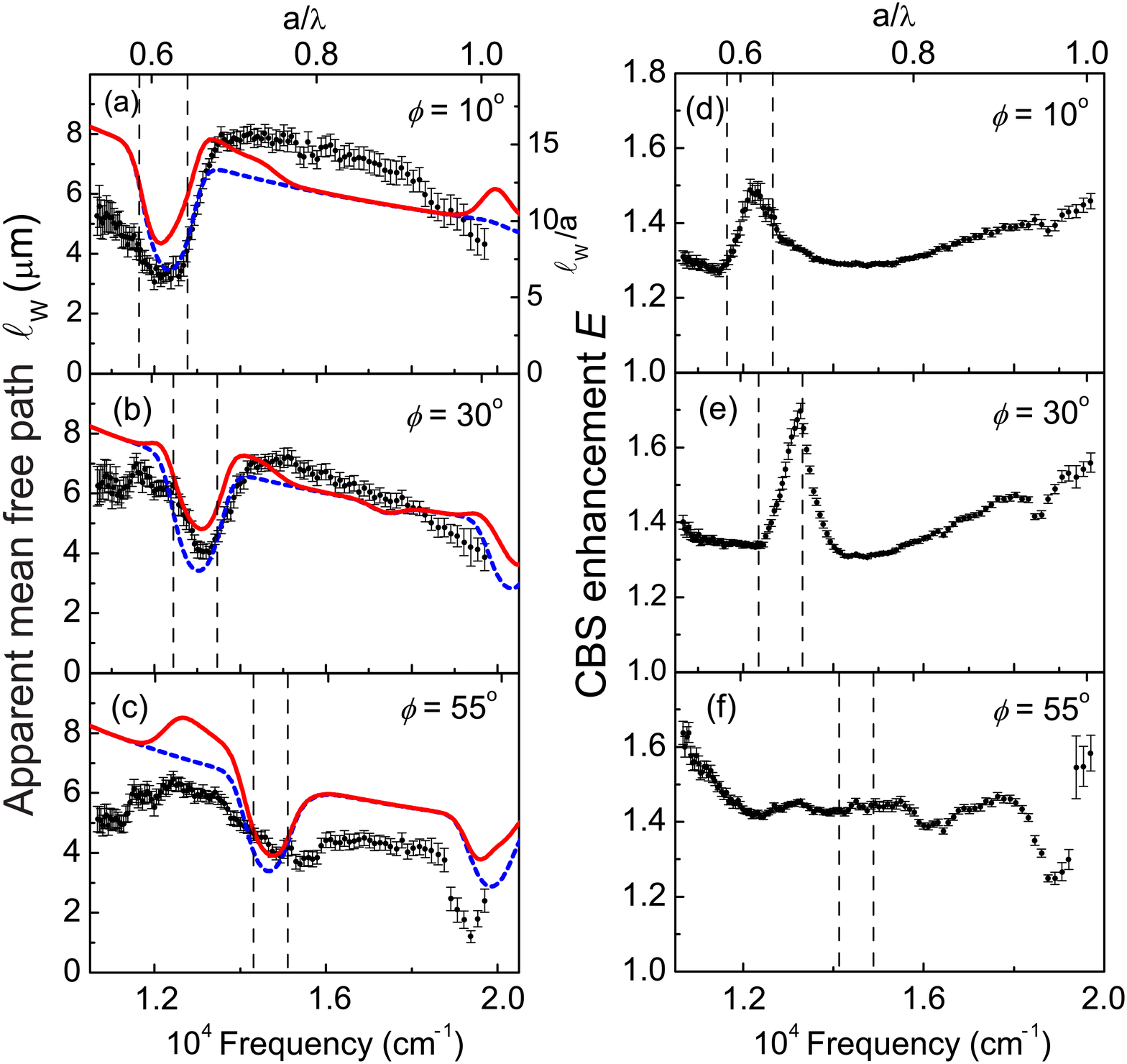}
\caption{\label{fig:cu180CBSfits} (a-f) Values of $\ell_W$ (a-c) and $E$ (d-f)
obtained from fits of CBS spectra for the 180-nm opal. Error bars denote 95\%
confidence interval of the fits. Lines: calculations using Eq.~(\ref{eq:ellW}) with (line, red) and without (dash, blue) internal reflection correction.}
\end{figure}

The CBS enhancement factors in Fig.~\ref{fig:CBSimage}(d) are
consistently low compared to values obtained for strongly
scattering random photonic media measured using the same
instrument.\cite{muskensOE08} This behavior is particularly
striking below the stop band, where photonic crystals are commonly
expected to behave as an effective
medium.\cite{Datta93,Genereux01} The enhancement factor ranges
between 1.4 and 1.5, which is larger than values reported by Huang
et al.\cite{HuangPRL01} but smaller than measured in
Ref.~[\onlinecite{KoenderinkPLA}]. A reduction of the enhancement
factor below the stop band was also reported in
Ref.~\onlinecite{Baumberg09}. The ideal enhancement factor of two
is expected in any multiple scattering sample, whether absorbing
and of finite size or not, as long as the sample obeys
reciprocity. However, if scattering components contribute that
have no distinct time reversed counterpart, such as single
scattering, the enhancement factor is reduced.\cite{Mark88} In
addition to this general trend, we find pronounced variations of
$E$ around the stop gap. At the band edge, two sharp troughs
appear in Fig.~\ref{fig:CBSimage}(d), which will be discussed
further below.

Results obtained from CBS spectra for the $r=180$~nm opal at
higher reduced frequencies are shown in
Fig.~\ref{fig:cu180CBSfits}. The spectra for the 180-nm opal yield
more information on the behavior above the L-gap, covering values
of $a/\lambda$ between 0.55 and 1.03. Values of $\ell_W/a$ are
around a factor three smaller for this opal than for the
$r=130$~nm sample, which is indicative of an increased disorder.
Apart from photonic band structure effects, we observe a gradual
decrease of $\ell \propto \omega^{-\alpha}$, which can be modeled
using $\alpha=0.75\pm 0.08$ for the spectrum at $\phi=55^\circ$.
The power-law exponent is much lower than that observed in
transmission measurements, emphasizing the nonuniversality of the
mean free path in opal photonic crystals, as opposed to earlier
suggestions.\cite{KoenderinkPRB05}

\begin{figure}[t] \newpage
\includegraphics[width=8.6cm]{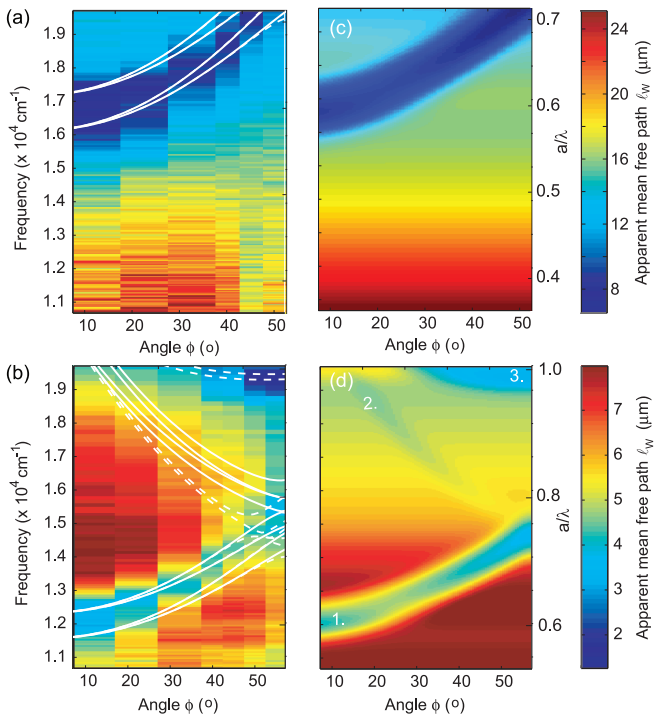}
\caption{\label{fig:allfitcmaps} Color online. (a) (Left) Color density maps of
$\ell_W$ obtained from fits to CBS cones for the $r=130$~nm opal.
(Right) Calculated $\ell_W$ using  Eq.~(\ref{eq:ellW}). (b) Same for
the $r=180$~nm opal. Lines indicate band structure from
Fig.~\ref{fig:reflspectra}. (c,d) Calculations made using
Eq.~(\ref{eq:ellW}). Numbers indicate contributions from three bands
of Eq.~(\ref{eq:3-bandrefl}).}
\end{figure}

Similar to the 130-nm opal, a distinct trough in $\ell_W$ is
observed in Fig.~\ref{fig:cu180CBSfits} which shifts with the
frequency of the stop gap for increasing angle of incidence. From
the width of the stop gap, we estimate a Bragg length $L_B$ of
around 5~$\mu$m, indicating a ratio $\ell/L_B$ of $\sim 1$ for
this more strongly disordered opal. We point out the asymmetry of
both the mean free path and the enhancement factor over the stop
band in Fig.~\ref{fig:cu180CBSfits}(b). Such asymmetry cannot be
explained by the group velocity behavior as this is symmetrical
around the stop gap.\cite{Garcia09}

At angles of incidence above $45^\circ$ [cf.
Fig.~\ref{fig:cu180CBSfits}(c)], a second trough in $\ell_W$
appears around 19000~cm$^{-1}$. Comparison with the band structure
calculations of Fig.~\ref{fig:reflectivities} reveals that this
feature can be attributed to another Bragg reflection at oblique
$(11\overline{1})$ planes, which appears around  $a/\lambda \sim
1$ at the $U$ and $X$ points in the Brillouin zone. The presence
of this band is also clearly observed as a trough in the
enhancement factor $E$ in Fig.~\ref{fig:cu180CBSfits}(f). The
presence of this higher-order diffraction peak demonstrates that
the observed behavior has to be described by the full 3D photonic
band structure and not just by a simple one-dimensional Bragg
reflector.

\begin{figure}[t] \newpage
\includegraphics[width=8.6cm]{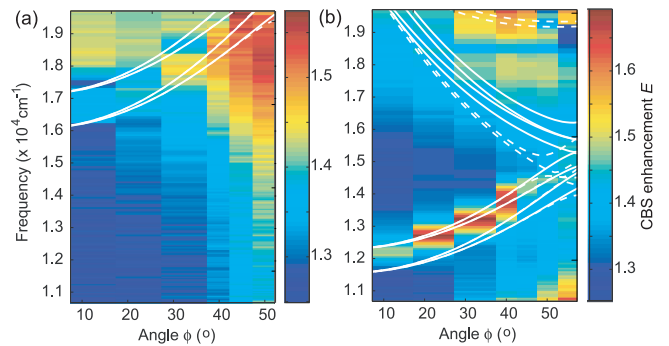}
\caption{\label{fig:enhancementmaps} Color online. Fitted CBS enhancement factor
for the $r=130$~nm opal (a) and the $r=180$~nm opal (b). Lines
indicate band structure from Fig.~\ref{fig:reflspectra}.}
\end{figure}

The extensive information on the diffuse scattering parameters
obtained from CBS spectra taken at different angles and
frequencies can be combined into maps of fitted values for
$\ell_W$ and $E$ as shown in Figs.~\ref{fig:allfitcmaps} and
\ref{fig:enhancementmaps}. Figures~\ref{fig:allfitcmaps}(a) and
(b) show the experimental data, while
Figures~\ref{fig:allfitcmaps}(c) and (d) are results from
calculations discussed below in Sec.~\ref{sec:calculations}. The
CBS enhancement $E$ is shown in Fig.~\ref{fig:enhancementmaps}. In
the following section, we discuss the different contributions to
the wavelength dependence of $\ell_W$ and $E$ in more detail.

\section{Model calculation of band structure effects on angle-dependent CBS cone width}
\label{sec:calculations}
\subsection{General model for CBS-cone width}
In this section we discuss the combined frequency and
angle-dependence of the width of the CBS cone. Our starting point
for the description of diffuse light scattering in the crystal is
the theory for random photonic media that we extend for the
effects of the photonic crystal band
structure.\cite{KoenderinkPLA,KoenderinkPRL03} According to
predictions by Ref. [\onlinecite{KoenderinkPLA}], the cone width
is modified by two effects: internal
reflection\cite{KoenderinkJOSAB} and the limited penetration depth
of the incident light when its wavelength matches the Bragg
condition. A larger internal reflection results in longer paths,
providing a narrowing of the CBS cone. In contrast, reduction of
the penetration depth leads to shorter paths as the light has a
larger probability to exit via the surface, resulting in a
broadening of the cone. It has been shown that the resulting
correction to the CBS cone can be included in the random diffusion
model by introducing an apparent mean free path $\ell_W$ which is
given by\cite{KoenderinkPLA}
\begin{equation}\label{eq:ellW}
\ell/\ell_{W}=\frac{(1+\tau_0)^2}{1+2\tau_0}\frac{\xi
[1+2(\epsilon + \tau_e)\xi]}{[1+(\epsilon+\tau_e)\xi]},
\end{equation}
with
\begin{equation}
\epsilon\equiv \frac{\overline{R}}{1-\overline{R}} \, .
\end{equation}
Here $\overline{R}$ is the internal reflection coefficient
integrated over all angles as put forward in the theory for
multiple light scattering.\cite{Lagendijk89,Zhu91} The penetration
depth of the incident light is included in the parameter $\xi
\equiv 1+\ell/L_B$. The Bragg length varies between infinity (no
Bragg diffraction) and $2d_{111}/\pi S$, with $S=0.06$ the
photonic strength and $d_{111}$ the lattice constant in the
(111)-direction. We use the extrapolation length ratios
$\tau_0=2/3$ and $\tau_e=\tau_0(1+\overline{R})/(1-\overline{R})$
that were originally derived for random
media.\cite{Lagendijk89,Zhu91}

\begin{figure}[t] 
\includegraphics[width=6.8cm]{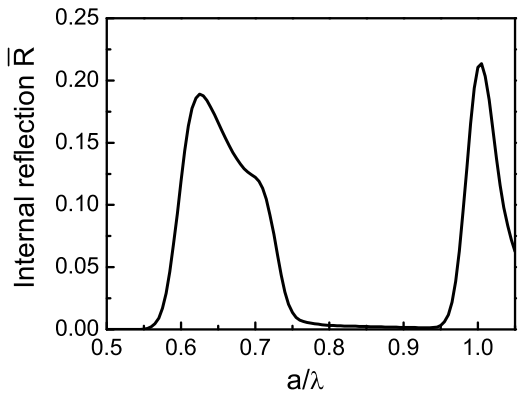}
\caption{\label{fig:intrefl} Internal reflection coefficient
calculated using the three-gap model of
Eq.~(\ref{eq:3-bandrefl}).}
\end{figure}

\subsection{Internal reflection}
We propose to capture the essential optical properties of the
photonic crystals by the three dominant stop gaps. Therefore, the
internal-reflection coefficient of the opals is modeled as the sum
of three Gaussian reflection peaks
\begin{equation}\label{eq:3-bandrefl}
R^D_\omega(\mu_i) =
\sum_{j=1}^3 R_j(\mu_i) \exp\left[-
\frac{(\omega-\omega_j(\mu_i))^2}{2\Delta
\omega_j(\mu_i)^2}\right] \, ,
\end{equation}
with angle-dependent peak reflectivities and widths $R_{j}(\mu_i)$
and $\Delta_{j}(\mu_i)$, with $\mu_i\equiv \cos \phi$. Following
earlier successful descriptions of the angle dependent
reflectivities \cite{VanDiel2000} and escape
function\cite{KoenderinkPRL03}, we use the gaps along the LU
high-symmetry direction to model the reflectivity, and assume a
dependence on angle according to $R_j(\mu_i)=R_j \cos \phi +
R_{j,{\rm bg}}$, where the cosine-dependence is based on simple
Bragg diffraction. The frequencies $\omega_{1,2,3}$ are obtained
from the calculated band edges in
Fig.~\ref{fig:reflectivities}(d).

\begin{figure*}[t] \newpage
\includegraphics[width=16.0cm]{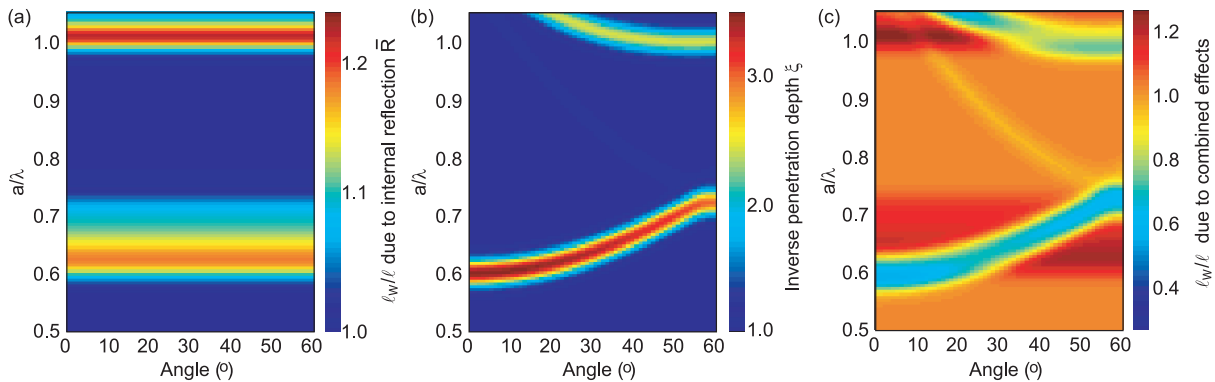}
\caption{\label{fig:modelbroadening} Color online. (a) CBS cone narrowing factor
$\ell_W/\ell$ calculated using Eq.~(\ref{eq:ellW}) including effect
of internal reflection $\overline{R}$ from Fig.~\ref{fig:intrefl}.
(b) Calculated penetration depth parameter $\xi$, and (c) CBS cone
narrowing resulting from combined effects of $\overline{R}$ and
$\xi$.}
\end{figure*}

Using the three-gap model described by Eq.~(\ref{eq:3-bandrefl}),
the internal reflection $\overline{R}$ is obtained by
angle-averaging for every frequency using
\cite{Lagendijk89,Zhu91,Durian96}
\begin{equation} \label{eq:intrefl1}
\bar{R}_\omega^D  =  \frac{3C_{2,\omega}  + 2 C_{1,\omega} }{3
C_{2,\omega} -2 C_{1,\omega}+2}
\end{equation}
with
\begin{equation}\label{eq:intrefl2}
C_{n,{\omega}}=\int_0^{\pi/2} R_{\omega}^D(\alpha)\,\cos^n(\alpha)
\sin\alpha\, d \alpha.
\end{equation}
The resulting values of the internal reflection for an opal
photonic crystal are shown in Fig.~\ref{fig:intrefl}. The internal
reflection due to photonic band structure ranges between zero and
14\% in the frequency range of interest. We notice three peaks in
the spectrum of $\overline{R}$, which can be attributed to the
L-gap and the avoided crossing at the U and W points. The internal
reflection contribution should not be confused with effects due to
density of states, which also are independent of angle. There is a
fundamental physical difference between these two physical
parameters: the internal reflection coefficient is the result of
gaps, in other words, of the absence of states. Conversely, the
density of states is a sum over states, and is thus the result of
the presence of states. Moreover, the variations in $\overline{R}$
are much greater than those in the density of states, see
Fig.~\ref{fig:reflspectra}(b) and Refs.~[\onlinecite{Nikolaev08},
\onlinecite{Li01}].

\subsection{Modification of diffuse source by Bragg diffraction}
At the center of the stop gap, the following expression holds for
the Bragg length $L_B$ in a two-band model\cite{Shung}
\begin{equation} \label{eq:Lbragg}
L_B=\frac{2d_{111}}{\pi S}
\end{equation}
The Bragg length varies between Eq.~(\ref{eq:Lbragg}) at the Bragg
condition and infinity outside the stop band. In absence of a
theoretical description of $L_B$, we assume here $L_B$ to be
inversely proportional to the reflectivity around the stop gap
\begin{equation}\label{eq:Lbraggnew}
L_B(\mu_i,\omega) =\frac{2d_{111}}{\pi S}
\frac{1}{R^D_\omega(\mu_i)} \, ,
\end{equation}
where $R^D_\omega(\mu_i)$ is the internal reflection coefficient
defined in Eq.~(\ref{eq:3-bandrefl}) and $S=0.06$ is the photonic
strength obtained from band structure calculations and
experiments. This definition of $L_B$ is chosen as it provides the
right value in the middle of the gap, the peak in $1/L_B$ follows
the reflection coefficient as expected, and outside the reflection
peaks $L_B$ becomes very large, as expected. Furthermore, a
reduced reflection coefficient yields an increase of $L_B$, which
is qualitatively consistent with experiments by Vlasov et al. and
Bertone et al.\cite{Vlasov99,Bertone99} More accurate modelling of
$L_B$ is beyond the scope of our semi-empirical analysis.

The modification of the diffuse source by Bragg reflection is
included in Eq.~(\ref{eq:ellW}) through the parameter
$\xi=1+\ell/L_B$. The parameter $\xi$ can be identified as the
inverse penetration depth of the light in the crystal due to Bragg
reflection.\cite{KoenderinkPLA} Below we will calculate the effect
of photonic stop bands on $\xi$ for the opals under study. We note
that the model presented for internal reflection here is a
rigorous framework in multiple scattering theory, barring the
introduction of a semi-empirical description of the band structure
in Eq.~(\ref{eq:3-bandrefl}) and Eq.~(\ref{eq:Lbraggnew}). These
approximations limit detailed comparison particularly at the edges
of the stop band. Our model is highly constrained, as the only
adjustable parameters that we allow to vary are the $R_{1,2,3}$
that are constant or very smoothly varying.

\subsection{Comparison with experiments}
\label{sec:comparison}

Using the model above, we will estimate the effects of the
internal reflection and source distributions on the CBS cone
width. Values of $\ell$, $R_1$, and $R_2$ were optimized to
achieve good agreement with the experimental data of
Fig.~\ref{fig:allfitcmaps}, yielding $R_1=0.5+0.2 \cos\phi$,
$R_2=0.02$, and $R_3=0.4$. These four coefficients together with
the overall scaling of the mean free path are the only adjustable
parameters used to fit the data for all frequencies and incidence
angles. The small value of $R_2$ is chosen since this band does
not have a pronounced appearance in our experimental data.
Figure~\ref{fig:modelbroadening}(a) shows the effect of internal
reflection only on $\ell_W$. The increase of $\overline{R}$
results in a narrowing of the cone and thus an increase of
$\ell_W$ according to Eq.~(\ref{eq:ellW}), which does not depend
on angle of incidence $\phi$. In contrast, the inverse penetration
depth $\xi$, shown in Fig.~\ref{fig:modelbroadening}(b), follows
the band structure according to Eq.~(\ref{eq:3-bandrefl}).

The two contributions are combined to yield the apparent mean free
path as shown in Fig.~\ref{fig:allfitcmaps}(c,d) for the two
opals. For comparison with the experimental data, we included the
power law dependence of the background mean free path in absence
of photonic effects. The calculated maps for $\ell_W$ in
Fig.~\ref{fig:allfitcmaps} agree well with most of the
experimental observations. The cross sections, shown as red lines
in Figs.~\ref{fig:CBSimage} and \ref{fig:cu180CBSfits}, reveal
that the depth of the correction on $\ell_W$ closely matches the
experimental behavior. However, the effect of internal reflection,
predicted as an increase of $\ell_W$ before the stop band at
larger angles, is not consistent with our experimental data. We
note that effects of internal reflection are generally difficult
to disentangle from smoothly varying corrections on $\ell_W$ of
the order of 10\%. The predominant modification of the cone width
in the photonic crystals under study is caused by the reduction of
the penetration depth, as is illustrated by calculations including
only this correction (dashed lines, blue).

\subsection{Enhancement factor}

Here, we will briefly discuss the remarkable behavior of the
CBS-enhancement factor. After elimination of spurious background
contributions to the intensity, variations in $E$ can only be
caused by changes in the intensity and polarization of
single-scattered light. The low enhancements at small reduced
frequencies are particularly puzzling, since photonic crystals in
the long wavelength limit are widely expected to behave as
effective weakly scattering media.\cite{Datta93,Vos96,Genereux01}
A similar reduction of $E$ in the long-wavelength limit was
observed for thin slabs of nanowires, which was explained by the
finite thickness $L$ of the layer of the order $L/\ell<10$ and the
concomitant suppression of higher scattering
orders.\cite{muskensNL09} For the current samples, in the long
wavelength regime, $L/\ell$ is still larger than 15. Therefore we
can exclude finite sample thickness as an origin for the low
enhancement.

The effect of stop gaps on single scattering may be caused by
changes in intensity or polarization of the single-scattered
light. While absorption loss does not change the enhancement
factor $E=2$ for an ideal experiment where only multiply scattered
light is collected, absorption might influence $E$ via the ratio
between single and multiply scattered light. However, absorption
would also be apparent as a cone rounding. By analyzing the cone
rounding, we note that the diffuse absorption length is equal to
or even greater than the sample thickness, i.e. 200~$\mu$m. This
value corresponds to an absorption decay length of 6 mm, in
agreement with our previous measurements.\cite{Vos01} Therefore,
we conclude that the role of absorption loss is negligible.

\begin{figure}[t] \newpage
\includegraphics[width=8.6cm]{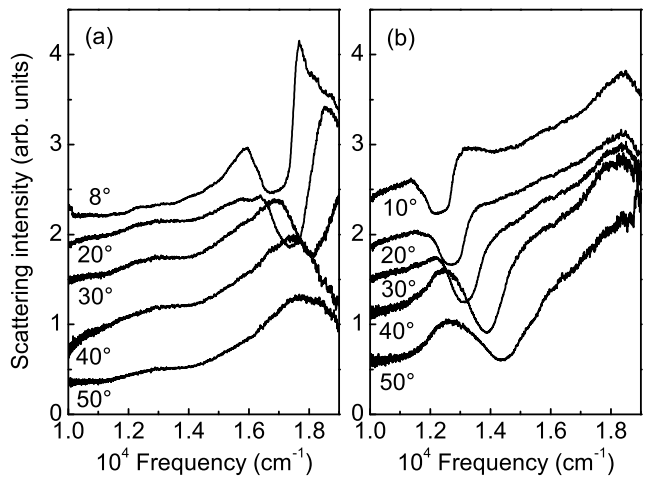}
\caption{\label{fig:scattering} Scattering intensity from the
130-nm (a) and 180-nm (b) opals, for angles of incidence from
8$^\circ$ to 50$^\circ$, measured at a detection angle of
5$^\circ$ using helicity conserving channel. Subsequent spectra
are offset by 0.5 for visibility, starting from $\phi=50^\circ$
(no offset).}
\end{figure}

Variations in the total scattering intensity are observed around
the stop gap in our samples, as shown in
Fig.~\ref{fig:scattering}, consistent with the behavior reported
in other work.\cite{Astratov2002,Baumberg09} We see a minimum in
the stop gap with two wings of high scattering at the band edges.
Inside the stop gap, the Bragg reflection removes intensity from
the diffuse scattering. This redistribution does not necessarily
influence the CBS enhancement factor considering that the Bragg
peak does not lead to additional single scattering. A decrease of
the incoherent single scattering background in the stop gap may
even be expected as single scattering from the photonic crystal
coherently adds to the specular Bragg reflection and is thus
suppressed at other angles. The overall increase of scattering
intensity toward shorter wavelengths and larger angles of
incidence is consistent with an increase of the enhancement factor
which can be attributed to more efficient randomization of light
in the medium.

Another pronounced feature in $E$ are the troughs at the band
edges observed for the $r=130$~nm opal in
Fig.~\ref{fig:CBSimage}(d). The conditions of these throughs
appears correlated with presence of sharp peaks in the scattering
intensity of Fig.~\ref{fig:scattering}(a), which currently is not
explained by any models. These effects resemble the characteristic
behavior of the group velocity as well as recent measurements of
scattering mean free path.\cite{Garcia09} Such a reduction of the
single scattering length $\ell_s$ at the band edges is not
observed in the apparent mean free path $\ell_W$ of
Fig.~\ref{fig:CBSimage}(c), as $\ell_W$ is averaged over many
different directions in the crystal. We propose that the CBS
enhancement factor is inherently more sensitive to directional
band structure effects through its dependence on
single-scattering, and therefore changes of $\ell_s$ can result in
variations of $E$. The large variations in $E$ observed in our
experiment call for new theory for diffuse light in (partially)
ordered nanophotonic materials.

Some of the variations in the CBS enhancement cannot be associated
with the behavior of the total scattering intensity in
Fig.~\ref{fig:scattering}, but may be associated to the use of
circularly polarized light in our experiments. Markedly, there is
the drop in $E$ associated with the second stop band at
$\phi=55^\circ$. This opposite behavior to the increase of $E$ in
the L-gap points to a difference in helicity of single scattering,
possibly due to Brewster angle effects.\cite{Baryshev06} For the
L-gap at $\phi=0^\circ$, the parallel and perpendicular components
perceive exactly the same Bragg reflectivities and thus no change
of the polarization state of circularly polarized incident light
should occur. Strong polarization dependence of light transmitted
and reflected from opals was identified in several
studies.\cite{Galisteo2003,Baryshev06,Romanov2009} Polarization
effects have also been found in Ref.~\onlinecite{Baumberg09},
where in particular cross-polarized CBS cones were reported. Our
measurements of the cross-polarized channel only showed an
angle-independent response; no cross-polarized CBS cones were
detected.

\section{Conclusions}
We have investigated multiple scattering of light in
three-dimensional polystyrene opal photonic crystals using
white-light coherent backscattering spectroscopy. Our experiments
significantly extend on earlier measurements by covering
simultaneously a large frequency and angular range. The
measurements of the backscattering cone width at frequencies near
the stop bands confirm the model by Refs.
[\onlinecite{KoenderinkPLA, KoenderinkJOSAB}] which includes
internal reflection and the limited penetration depth of the
incident light. The semi-empirical model matches the experiments
without the need to include effects of density of states or group
velocity, such as were proposed in Ref.~[\onlinecite{Garcia09}].
In addition, we report photonic band structure contributions in
the CBS enhancement factor. We propose that the CBS enhancement
factor, being particularly sensitive to single scattering, thus
contains information on properties which are washed out in other
aspects of the multiple scattering CBS cone. A better
understanding of group velocity and other band structure effects
to the enhancement factor will require further theoretical
studies.

The presented experimental approach using white-light coherent
backscattering opens up studies on diffuse light transport in
novel materials. Future investigations may look toward stronger
scattering photonic crystals such as TiO$_2$ or silicon inverse
opals or silicon inverse woodpiles\cite{Huisman10}, for which
effects of the density of states could be observable.  Other
examples where the interplay of disorder - or reduced symmetry -
and band structure is expected to result in novel phenomena are
photonic quasicrystals\cite{Lederman09} and optical
graphene.\cite{Segev09,BeenakkerEPL09}

\section{Acknowledgements}
We thank A. Lagendijk for fruitful discussion and access to
experimental equipment. This work is part of the research program
of the "Stichting voor Fundamenteel Onderzoek der Materie (FOM)",
which is financially supported by the "Nederlandse Organisatie
voor Wetenschappelijk Onderzoek (NWO)". WLV thanks NWO-Vici and
Smartmix Memphis, AFK thanks NWO-Vidi for financial support. OLM
is supported by HEFCE through a SEPnet lectureship.

\end{document}